\documentclass[twocolumn,pra,aps,showpacs,reprint,nofootinbib]{revtex4-1} %,superscriptaddress nomes aos bois
\usepackage{amsmath,mathrsfs,amsbsy,color,graphicx,bm,amsthm,amsfonts}
\usepackage{times}
\usepackage{graphicx}
\usepackage{ulem}
\usepackage{braket}
\usepackage{dsfont}

\newcommand{\be}{\begin{equation}}
\newcommand{\ee}{\end{equation}}
\newcommand{\br}{\begin{eqnarray}}
\newcommand{\er}{\end{eqnarray}}
\newcommand{\tr}{\text{Tr}}
\newcommand{\prr}{\text{Pr($R_{1}|R_{2}$)}}
\newcommand{\pll}{\text{Pr($L_{1}|L_{2}$)}}
\newcommand{\prll}{\text{Pr($R_{1}|L_{2}$)}}
\newcommand{\plr}{\text{Pr($L_{1}|R_{2}$)}}

\newtheorem{myprop}{Proposition}
\newenvironment{rcases}
  {\left.\begin{aligned}}
  {\end{aligned}\right\rbrace}

\begin{document}
\title{Time as Consequence of Internal Coherence} 
\author{Leandro R. S. Mendes }
\email{leandrorsm@gmail.com}
\author{Diogo O. Soares-Pinto}
\email{dosp@usp.br}
\affiliation{Instituto de F\'isica de S\~ao Carlos, Universidade de S\~ao Paulo, CP 369, 13560-970, S\~ao Carlos, S\~ao Paulo, Brazil}

\begin{abstract}
Time has been an elusive concept to grasp. Although we do not yet understand it properly, there has been advances made in regards to how we can explain it. One such advance is the Page-Wootters mechanism. In this mechanism time is seen as an inaccessible coordinate and the apparently passage of time arises as a consequence of correlations between the subsystems of a global state. Here we propose a measure that captures the relational character of the mechanism, showing that the internal coherence is the necessary ingredient for the emergence of time in the Page-Wootters model. Also, we connect it to results in quantum thermodynamics, showing that it is directly related to the extractable work from quantum coherence.

\end{abstract}

\maketitle

\section{Introduction}
Although everyone could agree that time passes when questioned about the nature of time, if it is only a parameter or an observable, mixed answers would be given. Some (or perhaps most) would state that time is nothing more than a parameter that appears in Schr{\"o}dinger's equation and that it is representative of a classical clock on the wall of a laboratory \cite{pash}. Others would want to elevate time to an observable and put it on an equal footing to other quantities, such as position and momentum, in a similar way to that done in special relativity \cite{vac2}. It seems that if time really is an observable it is an inaccessible one. One solution for the seemingly inaccessibility of time was given by Page and Wootters \cite{paw}. They argued that time could not be observed because there may exist a superselection rule (SSR) for the energy, similar to there being a SSR for charge \cite{charge}. This statement leads to the following question: if there is an SSR for energy how do we agree that time passes? Page and Wootters (PaW) proposed that time emerges from correlations between subsystems in such a way that part or parts of the subsystem act as clocks for the rest, and in respect to which time flows. Today this is recognized as the PaW mechanism. Although the mechanism had been forgotten for some time because of criticisms \cite{ku}, a few ways to overcome these have since been presented \cite{gamb,glm,chia,dolby}, reigniting interest in the mechanism \cite{more,tobi,argen}.  

The system utilized to demonstrate that this mechanism is possible is composed of two non-interacting qubits, represented by spin-half particles prepared on the state $\ket{+} = (\ket{0}+\ket{1})/\sqrt{2}$. With these states, the PaW mechanism demonstrates that, even when there is a invariance relative to time, that is, the global state is a stationary state, the particles exhibit a certain type of time evolution that is seen through conditional probabilities. These probabilities reflect the direction in which the spin of each particle points given the other particles' spin direction. We can define the relative positions of the direction of the spins as follows: if the particles' spin is pointing it represents 12 o'clock and if the particles' spin points to the left then it represents 6 o'clock. Then the spin of the second particle can be deduced and, depending on the global state, the spin of the second particle can be deduced with very good accordance, even recovering the Schr{\"o}dinger's equation in regard to clock time. The connection is made through the total Hamiltonian $H=H_s+H_c$, and the extra requirement that the global state is an eigenstate of $H$ such that $H\ket{\Psi}=0$. Choosing $\ket{\phi(\tau_{0})}$ as the initial state of the clock, its evolution is governed by the clock Hamiltonian $H_{c}$ as $\ket{\phi(\tau)}= e^{-iH_{c}(\tau-\tau_{0})}\ket{\phi(\tau_{0})}$. It is shown that the state of the system will evolve according to clock time, by projecting the clock states onto the the global state \be \ket{\psi(\tau)}_{s}= e^{-iH_{s}(\tau-\tau_{0})}\ket{\psi(\tau_{0})}_{s},\ee where $H_s$ is the system Hamiltonian and $\ket{\psi(\tau)}_{s} = \braket{\phi(\tau)|\Psi}$. Then time can be seen as what is read on the clock, delivering a different type of clock model, which we are going to reference as the PaW clock.

Nevertheless, the study of quantum coherence has received a lot of attention lately because of its widely applicability in quantum technologies, which use purely quantum mechanical properties, and interesting phenomena that can be explained by it. Mainly, the advances are being achieved by using tools of quantum information theory, in the form of several frameworks for resource theories of coherence \cite{colo}. One of the initial proposals, made by Baumgratz, Cramer and Plenio \cite{bcp}, establishes a certain group of rules that any coherence measure has to obey to be considered as a proper monotone for coherence. One monotone for such a framework is the relative entropy of coherence $\min_{\tau \in \mathcal{I}} S(\rho||\tau)$, where $\tau$ belongs to the set of incoherent states $\mathcal{I}$. This measure admits a closed form \be\label{bc} C_r(\rho) = S(\Delta(\rho)) - S(\rho),\ee where $\Delta(\cdot)$ will be referred to as a \textit{fully dephasing operation}, represented by \be \Delta(\rho)= \sum^{d-1}_{i=0}\ket{i}\bra{i}\rho\ket{i}\bra{i},\ee where $d$ is the dimension of the Hilbert space $\mathcal{H}$ and $S(\rho) = -\tr(\rho \log \rho)$ is the von Neumann entropy.

 %Defining a map characterized by a set of Kraus operators \be \Lambda(\rho) = \sum_{n} K_{n}\rho K^{\dagger}_{n},\ee with the condition that if an incoherent state, $\tau \in \mathcal{I}$  that act upon by those operators remain incoherent for all $n$ (i.e. $K_{n}\mathcal{I}K^{\dagger}_{n} \subset \mathcal{I}$). It is important to note that all incoherent states are defined to be diagonal in respect to a given basis $\{\ket{i}\}$. 

Recently it was shown that coherence is a necessary ingredient to describe certain thermodynamic processes when considering the set of thermal operations, and is also relevant when connected to the study of quantum speed limits. Both connections required the same notion of coherence, where coherence is seen as a special case of asymmetry in the system relative to time translations \cite{matteo1,diego,marvian1}.

In Sec. II we explain the important aspects of the PaW model and introduce the Bell-diagonal states. In Sec. III we present the main result, a measure for the internal coherence and its connection to the PaW model. In Sec IV we briefly address the problem of the macroscopic case. Finally, in Sec. V and VI we connect the proposed measure to results in quantum thermodynamics.  
%Here we show that the relative entropy of coherence captures the split into internal and external coherence and propose a measure that represents the resource behind the Page-Wootters' model, being that the internal coherence of a state. 

%For the first case it is considered two systems representing a state and a bath, with Hamiltonians $H_A$ and $H_B$ respectively, that are put in thermal contact with one another. The set of thermal operations is defined as given by the operations that take the system to an state $\hat{\sigma}$ by means of a Stinespring dilation with the condition that any effect that causes an interaction must commute with the total Hamiltonian $H_A+H_B$. This set was shown to be a strict subset of the time translation invariant operations. For the second case the connection is faster to make, the measure of interest is the time that is going to take for a state $\rho_{t=0}$ to evolve to a distinct state $e^{-i H t}(\rho_{t=0})e^{i H t}$. It is straightforward to see that any state that is incoherent in the energy basis, that in this case takes a block diagonal form, is also going to be invariant under time translation, a condition required for any symmetric state. If a state is symmetric under time translations the speed of evolution of this states is zero, it never evolves. Therefore any quantifier of the asymmetry of a state in regards to time translations is a quantifier of the coherence and of the speed of evolution. 

\section{PaW revisited}

In the PaW model a world with an energy SSR is considered, which then implies on an inaccessible time coordinate \cite{barlet}. The inaccessible time coordinate imposes a restriction on the knowledge that we have about the states under study, in a way that our lack of knowledge of the external coordinate will reflect a lack of knowledge of any state in relation to it. This is represented in the model by the action of the operation \cite{marvian2} (with $\hbar = 1$) \be\label{deco} \mathcal{D}(\rho) = \frac{1}{T}\int^{T}_{0}{} e^{-i H_z t}\rho e^{i H_z t} dt, \ee where $T$ will be the period of the Hamiltonian used. This operation, which here is going to be called a \textit{dephasing operation}, is going to average out the action of the elements of the group of translations generated by the Hamiltonian $H_z$, which will result in a mixed state in our reference frame. The Hamiltonian for the PaW clock is the two-spin non-interacting Zeeman Hamiltonian \be H_z = -h \left(\sigma^{1}_z\otimes\mathds{1}^{2}+\mathds{1}^{1}\otimes \sigma^{2}_z\right),\ee where $\sigma^{i}_z$ is the Pauli-$z$ matrix for the $i$th qubit and $h$ is a constant representing the magnetic field. In terms of the density matrix, the action of the dephasing operation is to take a state $\rho$ to a state $\mathcal{D}(\rho)$ that is block-diagonal in the energy basis. 

As mentioned above, the model makes use of conditional probabilities as a way to show the correlation that arises between clock and system. Here we are going to adopt the ``right", or 12 o'clock, position being given by the state $\ket{+}$ and the ``left", or 6 o'clock, position being the state $\ket{-}$. As one of the qubits is going to act as the clock, we are going to label the other qubit as the system. We define the probability of agreement as given by the conditional probability that if one qubit is measured as pointing in one direction the other qubit will be pointing to the same direction, either clock right and system right or clock left and system left, as $\prr=\tr(\mathcal{D}(\rho)E_{R_{1}R_{2}})/\tr(\mathcal{D}(\rho)E_{R_{2}})$ and $\pll=\tr(\mathcal{D}(\rho)E_{L_{1}L_{2}})/\tr(\mathcal{D}(\rho)E_{L_{2}})$, respectively. $E_{R_{2}R_{2}}$ and $E_{L_{2}L_{2}}$ are the optimal projectors to distinguish between $\ket{+}$ and $\ket{-}$, belonging to the set of operators $\sum_{\eta, \mu} E_{\eta \mu }$, where each $\eta$ and $\mu$ represents a right or left. In the same manner the probabilities for opposite directions will be denoted by $\prll$ and $\plr$. \footnote{It should be noted that we only take probabilities of agreement into consideration for conciseness, a perfectly good clock could be built with the probabilities for opposite directions and the main result would not be altered.}

We consider two-qubit Bell-diagonal states \cite{bd} as initial states for the model. These states have maximally mixed reduced density operators and can be represented as 
\be\rho=\frac{1}{4}\left(\mathds{1} + \sum^{3}_{i=1}c_{i}\sigma_{i}\otimes\sigma_{i}\right),\ee where the parameters $c_{i}$, with $-1\leq c_{i} = \tr(\rho \sigma_{i} \otimes \sigma_{i}) \leq 1$, form a triplet that determines whose states are physically acceptable i.e., with non-negative eigenvalues, $\lambda_{\gamma \nu} \geq 0$. The eigenvalues can be obtained using \be \lambda_{\gamma \nu} = \frac{1}{4}\left[1+(-1)^{\gamma}c_1 - (-1)^{\gamma + \nu}c_2 + (-1)^{\nu}c_3\right],\ee where $\gamma$, $\nu$ = \{0,1\}. In the $\sigma_z$-basis they take an X form 
\be
  \rho=
  \frac{1}{4}\left[ {\begin{array}{cccc}
   1+c_3 & 0 & 0 & c_1-c_2 \\
   0 & 1-c_3 & c_1+c_2 & 0 \\
	 0 & c_1+c_2 & 1-c_3 & 0 \\
	 c_1-c_2 & 0 & 0 & 1+c_3 \\
  \end{array} } \right].
\ee Therefore the action of the dephasing operation results in \be
  \mathcal{D}(\rho)=
  \frac{1}{4}\left[ {\begin{array}{cccc}
   1+c_3 & 0 & 0 & 0 \\
   0 & 1-c_3 & c_1+c_2 & 0 \\
	 0 & c_1+c_2 & 1-c_3 & 0 \\
	 0 & 0 & 0 & 1+c_3 \\
  \end{array} } \right].
\ee

Since we are writing the Bell-diagonal states in the basis of the Hamiltonian, the probability of agreement for the two qubit Bell-diagonal state is \be \prr = \frac{1}{4}\left(2+c_1+c_2\right). \ee Thus we can study the impact on the probability of agreement of the two qubits for several states just controlling the parameters $\{c_1,c_2,c_3\}.$ A few interesting sets of parameters are those where: ($i$) $c_1+c_2 = 1 \ \text{with} \ c_3=0$, ($ii$) $c_1 = -c_2$ and $c_1=c_2=0$, both for any $c_3$, and ($iii$) $c_1=c_2=1$ and $c_3 = -1$. The first set is composed of all dephased states that have the same form as the ones studied in Ref. \cite{jw}. They act as a paradigm for the results. Therefore, we should be able to find the same probabilities; indeed, the probability of agreement, $\prr=0.75$, is the same as that found in Ref. \cite{jw}. The second set has conditional probabilities $\prr=\pll=\prll=\plr=0.5$, which tells us that there is an equal chance that, upon measuring the clock the system can be found in any of the two directions. It follows that there is no correlation between the clock and the system; in other words, there is no sense of time given by the conditional probabilities. This is because the flow of time in the model is represented as clock time. If the clock does not correlate with the position of the system  there can be no established causal connection, and we cannot say that the clock is measuring time. In the opposite direction the last set, which in fact is composed of only one element, given by one of the vertices of the tetrahedron formed by the Bell-diagonal states, gives a probability of agreement $\prr=1$, which is perfect agreement. The qubits are always going to be pointing in the same direction. 

It is very interesting to note the different outcomes in relation to the conditional probabilities for the parameters $\{1,1,-1\}$ and $\{1,-1,1\}$. Respectively, these parameters correspond to two Bell states: $\ket{\psi^{+}}=\left(\ket{01}+\ket{10}\right)/\sqrt{2}$ and $\ket{\phi^{+}}=\left(\ket{00}+\ket{11}\right)/\sqrt{2}$. These are two pure maximally entangled states, that result in two drastically different results. One gives the best possible probability  and the other the worst possible probability. This indicates that entanglement between the subsystems in the initial state, before the dephasing operation, is not responsible for the working of the PaW clock. Which does not indicate that entanglement of the dephased state is not required for the mechanism. In the next section we will see that in fact neither entanglement before nor entanglement after the dephasing operation can be connected to the emergence of time in the PaW model.  

\section{Internal coherence}

We can now present the main result of this work. The monotone presented in Eq. (\ref{bc}) measures how distinguishable a general state $\rho$ is from an incoherent state $\tau$. But the model described in the previous section clearly specify states which are symmetric in relation to the group of translations generated by $H_z$, or, in other words, block diagonal states. This implies that, inside the PaW universe, these are the only available states. If we start from the space containing all the density operators, what happens if we choose only those states that have a block diagonal form? The result is a measure of the distinguishability from the block diagonal state that is now representing our system and the closest incoherent state $\tau \in \mathcal{I}$ \be\min_{\tau \in \mathcal{I}}S(\mathcal{D}(\rho)||\tau),\ee where the dephasing operation is used to guarantee that the state is indeed block diagonal. 

It can be shown\footnote{See Appendix A.} that the relative entropy of coherence, when performing minimization from the block diagonal states to the incoherent states, is equivalent to Eq. (\ref{bc}) when we consider a dephased state, $\mathcal{D}(\rho)$, \be C_r(\mathcal{D}(\rho))=\min_{\tau \in \mathcal{I}}S(\mathcal{D}(\rho)||\tau).\ee This fact provides a very intuitive reasoning for the physical interpretation of such measures in terms of the coherences of a state. This result can be encapsulated in the following proposition. 

\begin{myprop}
The relative entropy of coherence defined in regards to the set of incoherent states can be broken down in terms of two types of coherences, the internal coherence and the external coherence, as \[C_r(\rho) = C_r(\mathcal{D}(\rho))+ A_G(\rho).\] Therefore the relative entropy of coherence is a measure of the total coherence, while $A_G(\rho)$ is a measure of the external coherence and $C_r(\mathcal{D}(\rho))$ is a measure of the internal coherence of a state.
\end{myprop} 

Here we take the same definition used in Ref. \cite{c-w}, so by internal coherence we mean the presence of off-diagonal elements of the density matrix of the state in the energy eigenbasis with the same energy. Then external coherence will be represented by off-diagonal elements with different energies. A demonstration of Proposition 1 can be found in Appendix B. The quantity $A_G(\rho)$, usually referred to as asymmetry or Holevo asymmetry, is a recognized measure of asymmetry which was first introduced in Ref. \cite{vac}, given by \be A_G(\rho) = S(\mathcal{D}(\rho)) - S(\rho). \ee When introduced, this measure was brought up in a very similar context to the PaW mechanism and used as a way to quantify the quality of a reference frame. As we saw, when asymmetry relative to time translation is invoked, it can be seen as coherence in the eigenbasis, therefore it is also a measure of coherence \cite{marvian2}. Since external coherence is always defined in regards to an external frame of reference, necessarily any measure capable of discerning the effects of said reference frame is going to be a measure for the external coherence. It follows then that the necessary coherence in any task where invariance over time translations is a factor (e.g., quantification of reference frames, quantum speed limits and quantum metrology) is the external coherence. 

We notice a separation in regards to the frameworks used to establish these measures. While $A_G(\rho)$ is a measure for \textit{unspeakable coherence} \cite{marvian2}, $C_r(\mathcal{D}(\rho))$ belongs to a class of incoherent preserving operations and cannot detect invariance over the dephasing operation. After all it is equivalent to how distinguishable a block diagonal state $\mathcal{D}(\rho)$ is from the closest incoherent state $\tau$. Therefore, it is only a measure of \textit{speakable coherence}. This reinforces the conclusion that this measure should be seen as a measure for relative phases between subsystems, that is, a measure for internal coherence. Given both definitions, it is straightforward to justify $C_r(\rho)$ as a quantifier for the total coherence.  

In terms of the PaW mechanism, the division of the total coherence into internal and external tells us which quantity is responsible for the PaW clock working. When applying the dephasing operation its action averages over the possible phases of the unknown time reference frame. Hence, this operation is going to eliminate any external coherence, if any, that the global state of the system has. The same cannot be said for the internal coherence---since it is defined between the subsystems as a relational degree of freedom, it is not erased even if we do not have access to a reference frame. Therefore, 

\begin{myprop}
The internal coherence, the relative entropy of coherence with minimization from a block diagonal state $\mathcal{D}(\rho)$ to the set of incoherent states $\mathcal{I}$, is responsible for the proper working of the PaW clock. And it is given by \[C_r(\mathcal{D}(\rho))= S(\Delta(\rho)) - S(\mathcal{D}(\rho)).\] with $C_r(\mathcal{D}(\rho))$ being the relative entropy of coherence applied to the dephasing operation $\mathcal{D}(\rho)$ and $\Delta(\rho)$ being the closest incoherent state to $\mathcal{D}(\rho)$.
\end{myprop} 
 
When evaluating this proposition it is important to understand the hypothesis taken in regards to the PaW clock. This mechanism is dependent on the conditional probabilities, as stated before. Thus we expect that these probabilities are going to give an indication of the performance of the clock in the model, especially in regards of its functioning. Even if the conditional probabilities do not describe all aspects of the model or the way in which we perform the measurements is not clarified, they are related to the principle by which the PaW clock works. Based on this principle it seems reasonable to say that for any state that renders a probability of agreement $\prr=0.5$, regardless of the subtleties of the process, when acquiring the information about clock time, the PaW clock will not work. In the same manner, if a state renders a probability of agreement $\prr=1$, we expect the clock to work near perfection. From this, it follows that any measure that is going to be necessary for the model to work (but not necessarily sufficient) must be zero when the clock does not work and maximum for the best clock. 

Granted this we can see that the internal coherence, as given by $C_r(\mathcal{D}(\rho))$, is necessary for the mechanism through direct calculation. For the Bell-diagonal states this measure admits an analytic form in terms of the triplet $\{c_1,c_2,c_3\}$ 

\begin{widetext}

\be C_r(\mathcal{D}(\rho))= -\frac{1-c_3}{2}\log(1-c_3) + \sum^{2}_{i=1}\frac{(1+(-1)^{i}c_1+(-1)^{i}c_2-c_3)}{4}\log(1+(-1)^{i}c_1+(-1)^{i}c_2-c_3).\ee

\end{widetext} 

When evaluated for the set with $c_1=-c_2$ and the set $c_1=c_2=0$, which corresponds to the sets with worst probabilities, this measure is always zero. For the case of perfect agreement, that is, for $c_1=c_2=1$ and $c_3=-1$, this measures is equal to 1. It turns out that this is not only the expected result, given our hypotheses, for the necessary measure for the PaW clock but also a very good indicator for the states which will improve the performance of the PaW clock. 

\begin{myprop}
For every family of parameters $c_3$ that is held constant, the internal coherence $C_r(\mathcal{D}(\rho))$ is greater for greater probability of agreement\footnote{This proposition is written with the convention that we adopted, which is that the desired probabilities are for agreement. As stated before a good clock can also be made with the convention for disagreement, in that case the with would read ``..$C_r(\mathcal{D}(\rho))$ is greater for greater probability of disagreement''.}.
\end{myprop}

By family we mean every group of states that are associated with a fixed $c_3$ for any $c_1$ and $c_2$ The only condition is that these parameters are within the range that returns physically acceptable density matrices for both the initial state $\rho$ and the dephased state $\mathcal{D}(\rho)$, with the latter being composed by a different set of values. Fixing the parameter $c_3$ in a given number, an order where all states with more internal coherence yielding best clock results, in the form of better conditional probabilities, emerges. It is worth noting that there is one family of states, for which $c_3=0$, were the agreement between probabilities and values for the internal coherence is in perfect accord with the best and worst results.  

We have seen that if the initial states have entanglement between the subsystems this did not affect the performance of the PaW clock, but does this remain valid if there is entanglement on the dephased state? After all it is also shown that the best result is obtained for the maximally entangled Bell state $\ket{\psi^{+}}$, which is also invariant under the dephasing operation, therefore we have a maximally entangled state as the dephased state. However the answer to this question is not necessarily. This can be seen by examining any state in the family that has $c_3=0$. We notice that there is no entanglement between the subsystems of the dephased state and yet this family returns a non-zero probability of agreement, that is, the PaW clock works without entanglement. Even in the original work, when dealing with density operators, and not wave functions, it is possible to see that the dephased state, which belongs to the same family that have $c_3=0$, is not entangled. This is shown to be the case in Appendix C. One could still claim that entanglement is necessary for the case of pure states. Although it may be true, it is hard to justify this reasoning given that the pure states that yield maximum probabilities of agreement are states with maximum internal coherence. It does not seem that one can take the internal coherence out of the picture and still be left with a working PaW clock. Based on these results it really seems that entanglement is neither a sufficient nor a necessary condition for the PaW clock to work. 

There is an interesting feature of treating the PaW mechanism as we did. Let the space of a quantum state be $\left(\mathbb{C}^{d}\right)^{\otimes N}$; this space carries representations of the unitary group $U(d)$ and the symmetric group $S_{N}$. By Schur-Weyl duality \cite{fu} there will be a decomposition of irreducible representations of these groups as \be \left(\mathbb{C}^{d}\right)^{\otimes N} \cong \bigoplus_{\lambda} V_{\lambda} \otimes P_{\lambda} \ee where the sum is taken over Young frames \cite{matth} and the spaces $\{V_{\lambda}\}$ and $\{P_{\lambda}\}$ are irreducible representations of the unitary and symmetric group respectively. It follows that the representation of each state is going to act irreducibly on their respective basis. For our case the dephasing operation in Eq. (\ref{deco}) will impose a U(1)-SSR, acting trivially on the elements of the symmetric subspace; therefore, the states on the space will not be perceived as invariant under permutations. For some elements changing the labels of the subsystems will change the global state, this can be translated to the possibility of distinguishing the system and the clock through the use of the symmetric subspace.

%which means that the dephasing operation will act trivially on the elements of the symmetric subspace, that in our case is $S_2$. Having no restriction on the }
\section{More than qubits}

An interesting approach to recover time from within quantum mechanics, %from the Page-Wootters' perspective for a macroscopic case
was done by Pegg \cite{pegg}. Here we briefly review this approach in the light of what we showed in the present work. Pegg started by considering a vector that would represent what is called the \textit{totality of physical reality}, where this vector was the zero energy eigenstate of the Hamiltonian of the Universe. Such a condition restricts the global state to being invariant under the dephasing operation, Eq. (\ref{deco}), a condition akin to that of the Page-Wootters model. The zero energy eigenstate was then shown to be proportional to $\sum^{s}_{j=0}\ket{\phi_j}$, where $s$ is related to the dimension, $s+1$, of the Hilbert space, and the states are connected through the Hamiltonian of the Universe \be \ket{\phi_m}= \exp[-iHm\beta]\ket{\phi_0},\ee for a constant $\beta$. The concept of time is then recovered by splitting each possible state of the Universe $\ket{\phi_m}$ as a system and a clock $\ket{\phi_m} = \ket{C_m}\ket{S_m}$. In this construction every possible state of the Universe is then associated with every time as measured internally by the clock. This approach is very similar in nature to what was expected by Wootters \cite{jw} for an $N$ particle universe. 

The conditions required for Pegg's mechanism were almost identical to the conditions that we used to draw the conclusion regarding the internal coherence relation to the emergence of time. Hence a very similar approach could be used to study this proposal: extend the possible states to encompass miced global states, project these to the zero energy eigenstate, choose at least a bipartition of the Universe to work as a clock, since there is no requirement set that the states corresponding to the system should be orthogonal to each other, and compute the probabilities. Owing to the generality of the relative entropy of coherence our proposed measure is not restricted to qubits---it can measure the internal coherence of a system with $d$ dimension (qudits), and could in principle be used to test the best global states for the model.

\section{Internal coherence and work}

Proposition 1 shows that the relative entropy of coherence does not always represent the same phenomenon. The physical interpretation of such a measure takes into account how distinguishable the initial state of the system is in regards to the final desired state. Therefore for an arbitrary state $\rho$ this ``distance" to the set of incoherent states reflects how close this arbitrary state is to being incoherent, which is interpreted here as a measure of total coherence. Without doubt, the set of incoherent states is further than the set of block diagonal states when taking a general state $\rho$, which is neither incoherent nor block diagonal. This is contained in Proposition 1 as \be C_r(\rho) \geq A_{G}(\rho),\ee following from the positivity of the relative entropy of coherence. Hence the relative entropy of coherence, representing the total coherence, is going to be an upper bound for the Holevo asymmetry, being also an upper bound for $C_r(\mathcal{D}(\rho))$, something expected for a measure of total coherence. This explains why the Holevo asymmetry is equivalent to the relative entropy, when a minimization is taken over all states that are invariant over a group action. Although any incoherent state is also going to be invariant over a group generated by the total Hamiltonian, the set of block diagonal states, is always closer to an arbitrary state $\rho$. This implies that the total coherence $C_r(\rho)$ given by how distinguishable a general state is from a group invariant state, is a completely different measure from $A_G$. 

%As an example, we can see that the results obtained in Ref. \cite{jan} \be\label{ja} F(\rho)-F(\mathcal{D}(\rho)) = kTA_G(\rho)\ee and in Ref. \cite{matteo1} \be F(\rho)-F(\Delta(\rho)) = kTC_r(\rho),\ee where the free energy is $F(\rho) = \tr(H\rho) - kTS(\rho)$, with $H$ being the total Hamiltonian and $T$ the temperature of the heat bath, are fundamentally different. While the definition given in Ref. \cite{jan} is equivalent to the work contained in the external coherence of state, that is evaluated as a difference of work with and without a frame of reference, the definition given in Ref. \cite{matteo1} using an operation that completely eliminates all coherence in a state that is equivalent to the fully dephasing operation $\Delta(\rho)$, is connected to the total coherence of the state which coincides with Eq. (\ref{ja}) in cases where there is only external coherence in the state.

To examine the relation between the proposed measure for internal coherence and extractable work from coherence let us consider the protocol proposed in \cite{paul,paul2}. These authors give a general protocol that could extract work from $n$ copies of an initial state $\hat{\rho}$ given access to a heat bath composed of an unlimited number of qubits on the thermal state \be\label{tau} \tau_B = \frac{e^{-\beta H_B}}{\mathcal{Z}},\ee where $H_B$ is the bath Hamiltonian and $\mathcal{Z}$ is the partition function $\mathcal{Z} = \tr\left(e^{-\beta H_B}\right)$. The process consists in applying a dephasing operation to all the $n$ copies of the state $\hat{\rho}$ yielding $n$ states $\mathcal{D}(\hat{\rho})$ which were then converted, individually, into thermal states $\tau_B$. The work produced in this process, which is shown to be equal to the difference in free energies from the initial states $\hat{\rho}$ and the thermal states in the limit that $n\rightarrow \infty$, could them be stored in a system with a weight that acted like a battery. Then in the single shot version, using a single copy of the state, the work that could be extracted is given by \be W_{Tot} = F(\mathcal{D}(\hat{\rho}))-F(\tau_B).\ee From this definition it follows that the total work that is extractable from the single shot regime of a state $\rho$ in transforming it to a fully dephased state is given by \be W(\rho) = F(\mathcal{D}(\rho))-F(\Delta(\rho)),\ee which can be easily demonstrated to be directly correlated with the internal coherence of the state \be\label{wo} W(\rho) = kTC_r(\mathcal{D}(\rho)).\ee  This result agrees with the interpretation given in \cite{c-w}, where the work extractable from coherence $W_{coh}$, which can be seen as a lower bound of the internal coherence $C_r(\mathcal{D}(\rho))$, is defined.

\section{Work locking and internal coherence}

It is very interesting to see the impact that internal coherence has on other frameworks where SSRs are at play. One example is the set of thermal operations. As a strict subset of the time covariant operations the set of thermal operations only includes those unitaries for which the conservation of energy is guaranteed, hence enforcing an energy SSR. In this set, allowed transformations, which take one state to another exhibit, a phenomenon called \textit{work locking} \cite{matteo2}. 

The setting is very similar to what was described above for the protocol that extracts a certain amount of work from a state. For a given initial state $\hat{\rho}$ and a bath $\rho_B$ we wish to perform the transformation \be \hat{\rho} \otimes \rho_B \rightarrow \hat{\sigma} \otimes \rho'_B, \ee with the difference being that the aforementioned protocol demands that energy is conserved on average, hence this gives a less strict set of operations where conservation of energy is not demanded at all times. The phenomenon appears when considering the work that can be extracted from the dephased version, $\mathcal{D}(\hat{\rho})$. When this is the case the transformation is given by \be \mathcal{D}(\hat{\rho}) \otimes \rho_B \rightarrow \mathcal{D}(\hat{\sigma}) \otimes \rho'_B, \ee which says that the work that is extractable from $\hat{\rho}$ is the same as that extractable from $\mathcal{D}(\hat{\rho})$, implying that the work from coherence is locked \cite{matteo2}. To ``unlock" this work an ancillary coherent state can be used. So given two states with coherence $\rho_1$ and $\rho_2$ for which \be
  \rho_1=\rho_2 = 
  \frac{1}{2}\left[ {\begin{array}{cc}
   1 & 1 \\
   1 & 1 \\
  \end{array} } \right],
\ee their dephased versions are going to output an incoherent $\mathcal{D}(\rho_1) = \mathcal{D}(\rho_2) = \tau_1$, and the work that can be extracted from $\tau_1$ is zero. Now if using as an initial state the product state $\rho_1 \otimes \rho_2$, considering one of the states as a coherent ancilla, the dephased state of the product is  \be
  \mathcal{D}(\rho_1 \otimes \rho_2) =
  \frac{1}{4}\left[ {\begin{array}{cccc}
   1 & 0 & 0 & 0 \\
   0 & 1 & 1 & 0 \\
	 0 & 1& 1& 0 \\
	 0& 0 & 0 & 1 \\
  \end{array} } \right],
\ee for which clearly $\mathcal{D}(\rho_1 \otimes \rho_2) \neq \tau_1 \otimes \tau_1$, making this state one that can be used to extract nonzero work from coherence. 

The role of internal coherence must be clear at this point, and also allows a different physical interpretation in regards to the role of the ancillary state used as a ``reference" to unlock the work from $\rho_1$. When dealing with individual systems $C_r(\mathcal{D}(\rho)) = 0$ and therefore $C_r(\rho)= A_G(\rho)$; the state representing a system will carry only extrinsic properties related to how it was produced. Hence, any coherence present in this individual state is assigned the role of external coherence, which cannot be done to the internal coherence as it is a relational property. Therefore, when working with thermal operations the information on this external coherence is lost, resulting in an incoherent state. To enable work extraction from coherence the ancilla is used, not as a reference for the initial state but as a way of generating a relative phase between subsystems. This phase can be viewed as a remnant of the extrinsic information carried by state and ancilla about the reference frame in which they were prepared, which allows a relational phase to be established, therefore generating internal coherence. That is why the work can be unlocked with as few as two states, which is the minimum necessary to generate internal coherence.

\section{Conclusion}	

We showed a split of the total coherence of a state into internal and external, as measured by the relative entropy of coherence. While the external coherence is directly connected to a framework for unspeakable information, and therefore connected to several tasks the use of asymmetry theory can be employed, the internal coherence seems to be only relative to speakable information, thus it is not useful for such tasks. Nevertheless, internal coherence seems to be a very important quantity. This was shown by investigating the PaW mechanism using two-qubit Bell-diagonal states. Considering the conditional probabilities as indicative of the performance and working of the PaW clock, we put forward evidence that internal coherence is a necessary quantity for the mechanism and not entanglement, as is usually believed. By investigating the parameters of the triplet of correlation $\{c_1,c_2,c_3\}$, we found that the lack of internal coherence in the system is always associated with the worst possible conditional probabilities of agreement between the clock and the system, and a maximum value of internal coherence is associated with the best possible probabilities of agreement. It was noted that internal coherence can also be used as a indicator for which states are going to give the best probabilities of agreement, given that a family of states related by the parameter $c_3$ is fixed, where greater internal coherence correlates with greater probability of agreement.

The internal coherence proved important also when dealing with quantum thermodynamics. In this case when the allowed operations and resources are carefully accounted for, in a way that no coherence can be sneaked in and that conservation of energy is always guaranteed, basically the same condition where an energy SSR takes place. The work that can be extracted in a single-shot regime is directly connected to the internal coherence between the states. This observation is useful when examining the phenomena of work locking, because it allows another interpretation of the use of additional copies to unlock the work of coherence. Instead of seeing the additional copies acting as frames of reference that provide an orientation which alleviates the constraints imposed by the SSR, adding copies to the initial state is interpreted here as a means to create internal coherence between the subsystems, which in turn can be used to extract work. 

This article focused on the Page-Wootters model when clock and system do not interact with each other. It is straightforward to consider what happens when this is not the case and the clock and system do see each other. A recent work \cite{meh} addressed this question, showing, that, when the interaction is present the state of the system obeys a time-nonlocal Schr{\"o}dinger equation. We aim to investigate this route in a future work.

\section{Acknowledgments}
The authors would like to thank Rafael S. do Carmo and Tiago Martinelli for critical reading of the manuscript and fruitful discussions. The project was funded by Brazilian funding agencies CNPq (Grants No. 142350/2017-6, 305201/2016-6), FAPESP (Grant No. 2017/03727-0) and the Brazilian National Institute of Science and Technology of Quantum Information (INCT/IQ).

\appendix
\section{Equivalence among measures}

Here we wish to show the validity of \be C_r(\mathcal{D}(\rho))=\min_{\tau \in \mathcal{I}}S(\mathcal{D}(\rho)||\tau).\ee The proof follow the same steps as known proofs connecting the Holevo asymmetry and the relative entropy of coherence when minimization over group invariant states is performed \cite{aberg,gour}. Then, from the relative entropy of coherence we have \begin{eqnarray} \min_{\tau \in \mathcal{I}}S(\mathcal{D}(\rho)||\tau) &=& \min_{\tau \in \mathcal{I}}\left\{ \tr[\mathcal{D}(\rho)\log \mathcal{D}(\rho)] - \tr[\mathcal{D}(\rho)\log \tau] \right\} \nonumber \\ &=& \tr[\mathcal{D}(\rho)\log \mathcal{D}(\rho)] - \max_{\tau \in \mathcal{I}}\tr[\Delta(\rho)\log \tau] \nonumber \\ &=& \tr[\mathcal{D}(\rho)\log \mathcal{D}(\rho)] - \tr[\Delta(\rho)\log \Delta(\rho)] \nonumber \\   &=& S(\Delta(\rho)) - S(\mathcal{D}(\rho)) \nonumber \\   &=& C_r(\mathcal{D}(\rho)),\end{eqnarray} where, in the second line we used the invariance under dephasing of $\tau$ and, in the third line we used the non-negativity of the relative entropy $S(\delta_1||\delta_2)\geq 0$.

\section{Demonstration of Proposition 1}

Recalling the definition for the closed form of the relative entropy of coherence \be C_r(\rho) = S(\Delta(\rho)) - S(\rho), \ee we can apply it to a dephased state $\mathcal{D}(\rho)$. This will yield \begin{eqnarray}\label{prop} C_{r}(\mathcal{D}(\rho)) &=& S(\mathcal{D}(\Delta(\rho))) - S(\mathcal{D}(\rho)) \nonumber \\ &=& S(\Delta(\rho)) -  S(\mathcal{D}(\rho)) \nonumber \\ &=& S(\Delta(\rho)) -  S(\mathcal{D}(\rho)) + S(\rho) - S(\rho) \nonumber \\ &=& C_r(\rho) - A_{G}(\rho), \end{eqnarray} where, in the second line, we used the invariance under dephasing of $\Delta(\rho)$, and the Holevo asymmetry is given by \be A_G(\rho) = S(\mathcal{D}(\rho)) - S(\rho). \ee Rearranging the result of Eq. (\ref{prop}) we get Proposition 1.

\section{Entanglement for the $c_3=0$ family}

To calculate the entanglement for this family we used the concurrence $\mathcal{C}(\rho)$ \cite{woo}, a widely known measure for entanglement. For two qubits it has an explicit form, \be \mathcal{C}(\rho) = \max \left\{0, \sqrt{\lambda_{1}} -\sqrt{\lambda_{2}} - \sqrt{\lambda_{3}} - \sqrt{\lambda_{4}}\right\}, \ee where each $\lambda_{i}$ is an eigenvalue of the matrix $\rho \tilde{\rho}$ in decreasing order and \be \tilde{\rho} = (\sigma_y\otimes \sigma_y)\rho^{*}(\sigma_y\otimes \sigma_y).\ee To show that for every state in the family $c_{3}=0$ has zero entanglement we first need to obtain the eigenvalues of $\rho \tilde{\rho}$. These are: \begin{eqnarray}\begin{rcases} \lambda_{1} &= \frac{c_{1}^{2}}{16} + \frac{c_{1}c_{2}}{8} - \frac{c_{1}c_{3}}{8} + \frac{c_{1}}{8} + \frac{c_{2}^{2}}{16} - \frac{c_{2}c_{3}}{8}& \\ &+ \frac{c_{2}}{8} + \frac{c_{3}^{2}}{16} - \frac{c_{3}}{8} + \frac{1}{16},& \\
\lambda_{2} &= \frac{c_{1}^{2}}{16} + \frac{c_{1}c_{2}}{8} + \frac{c_{1}c_{3}}{8} - \frac{c_{1}}{8} + \frac{c_{2}^{2}}{16} + \frac{c_{2}c_{3}}{8}& \\ &- \frac{c_{2}}{8} + \frac{c_{3}^{2}}{16} - \frac{c_{3}}{8} + \frac{1}{16},& \\
\lambda_{3} &= \lambda_{4} = \frac{c_{3}^{2}}{16} + \frac{c_{3}}{8} + \frac{1}{16}.& \end{rcases}\end{eqnarray} When setting $c_3=0$ we get \begin{eqnarray}\begin{rcases}  \lambda_{1} &= \frac{c_{1}^{2}}{16} + \frac{c_{1}c_{2}}{8} + \frac{c_{1}}{8} + \frac{c_{2}^{2}}{16} + \frac{c_{2}}{8}+& \frac{1}{16}, \\
\lambda_{2} &= \frac{c_{1}^{2}}{16} + \frac{c_{1}c_{2}}{8} - \frac{c_{1}}{8} + \frac{c_{2}^{2}}{16} - \frac{c_{2}}{8} +& \frac{1}{16},  \\
\lambda_{3} &= \lambda_{4} = \frac{1}{16}.& \end{rcases} \end{eqnarray}  Focusing on the first two eigenvalues, we see that they can be grouped as  \be \lambda_{1} = \frac{(c_{1} + c_{2})^{2}}{16} + \frac{(c_{1} + c_{2})}{8} + \frac{1}{16} \ee and
\be \lambda_{2} = \frac{(c_{1} + c_{2})^{2}}{16} - \frac{(c_{1} + c_{2})}{8} + \frac{1}{16}, \ee forming perfect squares. Therefore,\begin{eqnarray} \mathcal{C}(\rho) &=& \max\left\{0,\sqrt{\lambda_{1}} -\sqrt{\lambda_{2}} - \sqrt{\lambda_{3}} - \sqrt{\lambda_{4}}\right\} \nonumber \\ &=& \max\left\{0,\frac{c_{1}+c_{2} + 1}{4} - \frac{c_{1}+c_{2} - 1}{4} - \frac{1}{2}\right\}   \nonumber \\ &=&  \max\left\{0,\frac{1}{4}+ \frac{1}{4}- \frac{1}{2}\right\}  = 0.\end{eqnarray} 
 
%\vspace*{\fill}

\bibliographystyle{apsrev4-1}

\end{document}